\RequirePackage[2020-02-02]{latexrelease}
\documentclass[twocolumn]{revtex4}
\usepackage{graphicx}
\usepackage{dcolumn}
\usepackage{bm}
\usepackage{amsmath}
\usepackage{amsfonts}
\usepackage{amssymb}
\usepackage{color}
\usepackage{epstopdf}
\usepackage{float}
\usepackage{siunitx}
\providecommand{\U}[1]{\protect\rule{.1in}{.1in}}

\begin{document}
\title{Which is greater: $\boldsymbol{e^\pi}$ or $\boldsymbol{\pi^{e}}$? An unorthodox solution to a classic puzzle}
\author{Andr\'es Vallejo, Italo Bove}
\affiliation{\begin{small} Facultad de Ingenier\'{\i}a, Universidad 
de la Rep\'ublica, Montevideo, Uruguay\end{small}}
\date{\today}
\begin{abstract}
The question of the title is a famous puzzle in the field of recreational mathematics, 
and can be addressed by several approaches. A compilation of solutions, some of them very 
ingenious, can be found in \cite{Talwalkar}. In this contribution we present an alternative
solution based on the second law of thermodynamics. The method can be extended  
to derive a more general result involving the exponential function.
\end{abstract}

\maketitle

%
It is a well-known fact that when two bodies $A$ and $B$ at different temperatures 
are placed in contact, energy flows between them until thermal equilibrium is reached. 
Let us suppose that $A$ is an
incompressible solid with constant heat capacity $C$, initially at temperature $T_{1}=\pi$, and 
$B$ is a large thermal reservoir at temperature  $T_{B}=e$, where $e$ is Euler's number 
(both temperatures are measured in the same absolute temperature scale). We will 
assume that each of the systems only exchanges heat with the other.

Since there are no thermometers with infinite precision, the reader may argue that it is not 
possible to guarantee that the temperatures correspond precisely to these irrational numbers. 
However, for macroscopic systems with a large number of degrees of freedom, the temperature is 
usually considered as a continuous variable, that, a priori, can adopt any real value. In any case, 
if truncated expressions are employed, a single number after the comma is enough to obtain the 
correct relation between the exponentials in the title. 

Since the reservoir is large, the equilibrium is reached when the temperature of 
the solid equals that of the reservoir, which remains essentially constant. 
The entropy variation of the solid at the end of the process is:
	\begin{equation}\label{dS_S}
	\Delta S^{A}=C\log\left(\dfrac{T_{2}}{T_{1}}\right)=C(1-\log(\pi)).
	\end{equation}
The reservoir exchanges energy in an amount that is opposite to the 
solid's internal energy variation:
	\begin{equation}
	Q^{B}=-Q^{A}=-\Delta U^{A}=C(T_{1}-T_{2})=C(\pi-e),
	\end{equation}
so the entroy of the reservoir changes in the amount:
	\begin{equation}\label{dS_R}
	\Delta S^{B}=\dfrac{Q^{B}}{T^{B}}=C\left(\dfrac{\pi}{e}-1\right),
	\end{equation}
Finally, from Eqs. (\ref{dS_S}) and (\ref{dS_R}), we have that the global entropy change 
associated with the thermalization process is:
\begin{equation}\label{dS_U}
\Delta S^{Univ}=\Delta S^{A}+\Delta S^{B}=C\left[\dfrac{\pi}{e}-\log(\pi)\right]
\end{equation}

The key point is to note that, according to the second law of thermodynamics, the total entropy variation in any physical process must be non negative. Using that $C>0$, from Eq. (\ref{dS_U}) we obtain that
	\begin{equation}\label{ex}
	\dfrac{\pi}{e}-\log(\pi)\geq 0 \Longrightarrow \pi\geq\log(\pi^e)\Longrightarrow e^\pi\geq \pi^e,
	\end{equation}
so we conclude that $e^{\pi}$ is the greater of the two numbers. It is known as \textit{Gelfond's constant}, and it is a trascendental number whose approximate value is $e^{\pi}\simeq 23.14069$. On the other hand, $\pi^{e}\simeq 22.45915$, and it is unknown whether or not it is trascendental \cite{Pickover}.

We can formulate an alternative thermodynamic derivation of this result by considering a 
cylinder-piston device with diathermal walls containing a perfect gas in equilibrium in a
thermal bath. If the mass of the piston is such that the initial pressure in some unit 
system is $P_1=e$, and we suddenly increase the 
mass of the piston in such a way that the pressure exerted on the gas (in the same unit 
system) adopts the value $\pi$ \cite{Note1}, after transient oscillations, the system will 
reach a new equilibrium state with the bath, at pressure $P_2=\pi$ and at the initial 
temperature \cite{Mungan}. The interested reader may verify that the entropy analysis 
of this process leads to:
	\begin{equation}\label{DeltaS_univ}
	\Delta S^{Univ}=R\left[\dfrac{\pi}{e}-\log(\pi)\right],
	\end{equation}
from which the same conclusion is derived. 
It is also interesting to note that, if the final pressure (or, in the previous example, the initial temperature) is some arbitrary positive value $x$, an analogous reasoning allows to infer the following more general inequality:

	\begin{equation}\label{ineq1}
	e^x\geq x^e,\hspace{0.2cm}x\geq 0.
	\end{equation}

The possibility of deriving an inequality involving these famous mathematical 
constants by means of a physical law produces surprise and arouses interest in 
students. It usually motivates them to start exploring thermodynamic paths to generate 
other algebraic inequalities, which results in an improvement in students' skills in 
performing entropy analysis. 

Another interesting aspect that can be discussed in classroom is 
the logical status of this type of derivation. Since the results obtained are 
mathematical truths, and, therefore, independent of any physical considerations, 
it is interesting to discuss whether the thermodynamic derivation is, in fact, a proof  
in the mathematical sense. We refer the readers to Refs. \cite{Abriata,Deakin,Sidhu} 
to delve deeper into this issue. 

Instructors interested in this approach can find thermodynamic derivations 
of other famous inequalities in Refs. \cite{Cashwell,Landsberg,Tykodi} 
(inequalities between means), \cite{Plastino} (Jensen's inequality), or \cite{Vallejo} 
(Bernoulli's inequalities and bounds for the logarithmic function).


\section*{Acknowledgments}
This work was partially supported by Agencia Nacional de Investigaci\'on e Innovaci\'on and Programa de Desarrollo de las Ciencias B\'asicas (Uruguay). 


\end{document}